\documentclass[twocolumn,preprint]{aastex631}

\usepackage{amsmath,amstext,natbib,apjfonts,longtable,verbatim,float,graphicx,color,longtable}

\newcommand{\Ha}{\hbox{{\rm H}$\alpha$}}
\newcommand{\Hb}{\hbox{{\rm H}$\beta$}}
\newcommand{\Hg}{\hbox{{\rm H}$\gamma$}}

\newcommand{\HII}{\hbox{{\rm H}\kern 0.1em{\sc ii}}}
\newcommand{\HeI}{\hbox{{\rm He}\kern 0.1em{\sc i}}}
\newcommand{\HeII}{\hbox{{\rm He}\kern 0.1em{\sc ii}}}
\newcommand{\CIII}{\hbox{{\rm C}\kern 0.1em{\sc iii}{\rm ]}}}
\newcommand{\CIV}{\hbox{{\rm C}\kern 0.1em{\sc iv}}}
\newcommand{\NII}{\hbox{{\rm [N}\kern 0.1em{\sc ii}{\rm ]}}}
\newcommand{\OII}{\hbox{{\rm [O}\kern 0.1em{\sc ii}{\rm ]}}}
\newcommand{\OIII}{\hbox{{\rm [O}\kern 0.1em{\sc iii}{\rm ]}}}
\newcommand{\NeIII}{\hbox{{\rm [Ne}\kern 0.1em{\sc iii}{\rm ]}}}
\newcommand{\NeIV}{\hbox{{\rm [Ne}\kern 0.1em{\sc iv}{\rm ]}}}
\newcommand{\NeV}{\hbox{{\rm [Ne}\kern 0.1em{\sc v}{\rm ]}}}
\newcommand{\SII}{\hbox{{\rm [S}\kern 0.1em{\sc ii}{\rm ]}}}
\newcommand{\hst}{\textit{HST}}
\newcommand{\jwst}{\textit{JWST}}

\shorttitle{High-$z$ Emission-Line Ratios with JWST}
\shortauthors{Trump et al.}

\begin{document}

\title{\large \bf The Physical Conditions of Emission-Line Galaxies at Cosmic Dawn \\
from \textit{JWST}/NIRSpec Spectroscopy in the SMACS~0723 Early Release Observations}

\author[0000-0002-1410-0470]{Jonathan R. Trump}
\affiliation{Department of Physics, 196 Auditorium Road, Unit 3046, University of Connecticut, Storrs, CT 06269}


\author[0000-0002-7959-8783]{Pablo Arrabal Haro}
\affiliation{NSF's National Optical-Infrared Astronomy Research Laboratory, 950 N. Cherry Ave., Tucson, AZ 85719, USA}

\author[0000-0002-6386-7299]{Raymond C. Simons}
\affil{Space Telescope Science Institute, 3700 San Martin Drive, Baltimore, MD, 21218 USA}

\author[0000-0001-8534-7502]{Bren E. Backhaus}
\affiliation{Department of Physics, 196 Auditorium Road, Unit 3046, University of Connecticut, Storrs, CT 06269}

\author[0000-0001-5758-1000]{Ricardo O. Amor\'{i}n}
\affiliation{Instituto de Investigaci\'{o}n Multidisciplinar en Ciencia y Tecnolog\'{i}a, Universidad de La Serena, Raul Bitr\'{a}n 1305, La Serena 2204000, Chile}
\affiliation{Departamento de Astronom\'{i}a, Universidad de La Serena, Av. Juan Cisternas 1200 Norte, La Serena 1720236, Chile}

\author[0000-0001-5414-5131]{Mark Dickinson}
\affiliation{NSF's National Optical-Infrared Astronomy Research Laboratory, 950 N. Cherry Ave., Tucson, AZ 85719, USA}

\author[0000-0003-0531-5450]{Vital Fern\'{a}ndez}
\affiliation{Instituto de Investigaci\'{o}n Multidisciplinar en Ciencia y Tecnolog\'{i}a, Universidad de La Serena, Raul Bitr\'{a}n 1305, La Serena 2204000, Chile}
\affiliation{Departamento de Astronom\'{i}a, Universidad de La Serena, Av. Juan Cisternas 1200 Norte, La Serena 1720236, Chile}

\author[0000-0001-7503-8482]{Casey Papovich}
\affiliation{Department of Physics and Astronomy, Texas A\&M University, College Station, TX, 77843-4242, USA}
\affiliation{George P.\ and Cynthia Woods Mitchell Institute for Fundamental Physics and Astronomy, Texas A\&M University, College Station, TX, 77843-4242, USA}

\author[0000-0003-0892-5203]{David C. Nicholls}
\affiliation{Research School of Astronomy and Astrophysics, Australian National University, Canberra, ACT 2600, Australia}

\author[0000-0001-8152-3943]{Lisa J. Kewley}
\affiliation{Research School of Astronomy and Astrophysics, Australian National University, Canberra, ACT 2600, Australia}
\affiliation{ARC Centre of Excellence for All Sky Astrophysics in 3 Dimensions (ASTRO 3D), Australia}

\author[0000-0001-6776-2550]{Samantha W. Brunker}
\affiliation{Department of Physics, 196 Auditorium Road, Unit 3046, University of Connecticut, Storrs, CT 06269}

\author[0000-0001-8483-603X]{John J. Salzer}
\affiliation{Department of Astronomy, Indiana University, 727 East Third Street, Bloomington, IN 47405, USA}

\author[0000-0003-3903-6935]{Stephen M.~Wilkins} %
\affiliation{Astronomy Centre, University of Sussex, Falmer, Brighton BN1 9QH, UK}
\affiliation{Institute of Space Sciences and Astronomy, University of Malta, Msida MSD 2080, Malta}


\author[0000-0001-9328-3991]{Omar Almaini}
\affiliation{School of Physics and Astronomy, University of Nottingham, University Park, Nottingham NG7 2RD, UK}

\author[0000-0002-9921-9218]{Micaela B. Bagley}
\affiliation{Department of Astronomy, The University of Texas at Austin, 2515 Speedway, Stop C1400, Austin, TX 78712, USA}

\author[0000-0002-4153-053X]{Danielle A. Berg}
\affiliation{Department of Astronomy, The University of Texas at Austin, 2515 Speedway, Stop C1400, Austin, TX 78712, USA}

\author[0000-0003-0883-2226]{Rachana Bhatawdekar}
\affiliation{European Space Agency, ESA/ESTEC, Keplerlaan 1, 2201 AZ Noordwijk, NL}

\author[0000-0003-0492-4924]{Laura Bisigello}
\affiliation{Dipartimento di Fisica e Astronomia "G.Galilei", Universit\'a di Padova, Via Marzolo 8, I-35131 Padova, Italy}
\affiliation{INAF--Osservatorio Astronomico di Padova, Vicolo dell'Osservatorio 5, I-35122, Padova, Italy}

\author[0000-0003-3441-903X]{V\'eronique Buat}
\affiliation{Aix Marseille Univ, CNRS, CNES, LAM Marseille, France}

\author[0000-0002-4193-2539]{Denis Burgarella}
\affiliation{Aix Marseille Univ, CNRS, CNES, LAM Marseille, France}

\author[0000-0003-2536-1614]{Antonello Calabr\`o}
\affiliation{INAF Osservatorio Astronomico di Roma, Via Frascati 33, 00078 Monteporzio Catone, Rome, Italy}

\author[0000-0002-0930-6466]{Caitlin M. Casey}
\affiliation{Department of Astronomy, The University of Texas at Austin, 2515 Speedway, Stop C1400, Austin, TX 78712, USA}

\author[0000-0003-0541-2891]{Laure Ciesla}
\affiliation{Aix Marseille Univ, CNRS, CNES, LAM Marseille, France}

\author[0000-0001-7151-009X]{Nikko J. Cleri}
\affiliation{Department of Physics and Astronomy, Texas A\&M University, College Station, TX, 77843-4242, USA}
\affiliation{George P.\ and Cynthia Woods Mitchell Institute for Fundamental Physics and Astronomy, Texas A\&M University, College Station, TX, 77843-4242, USA}

\author[0000-0002-6348-1900]{Justin W. Cole}
\affiliation{Department of Physics and Astronomy, Texas A\&M University, College Station, TX, 77843-4242, USA}
\affiliation{George P.\ and Cynthia Woods Mitchell Institute for Fundamental Physics and Astronomy, Texas A\&M University, College Station, TX, 77843-4242, USA}

\author[0000-0003-1371-6019]{M. C. Cooper}
\affiliation{Department of Physics \& Astronomy, University of California, Irvine, 4129 Reines Hall, Irvine, CA 92697, USA}

\author[0000-0002-3892-0190]{Asantha R. Cooray}
\affiliation{Department of Physics \& Astronomy, University of California, Irvine, 4129 Reines Hall, Irvine, CA 92697, USA}

\author[0000-0001-6820-0015]{Luca Costantin}
\affiliation{Centro de Astrobiolog\'{\i}a (CAB/CSIC-INTA), Ctra. de Ajalvir km 4, Torrej\'on de Ardoz, E-28850, Madrid, Spain}

\author[0000-0002-5009-512X]{Darren Croton}
\affiliation{Centre for Astrophysics \& Supercomputing, Swinburne University of Technology, Hawthorn, VIC 3122, Australia}
\affiliation{ARC Centre of Excellence for All Sky Astrophysics in 3 Dimensions (ASTRO 3D)}

\author[0000-0001-7113-2738]{Henry C. Ferguson}
\affiliation{Space Telescope Science Institute, 3700 San Martin Drive, Baltimore, MD, 21218 USA}

\author[0000-0001-8519-1130]{Steven L. Finkelstein}
\affiliation{Department of Astronomy, The University of Texas at Austin, 2515 Speedway, Stop C1400, Austin, TX 78712, USA}

\author[0000-0001-7201-5066]{Seiji Fujimoto}
\affiliation{
Cosmic Dawn Center (DAWN), Jagtvej 128, DK2200 Copenhagen N, Denmark
}
\affiliation{
Niels Bohr Institute, University of Copenhagen, Lyngbyvej 2, DK2100 Copenhagen \O, Denmark
}
\author[0000-0003-2098-9568]{Jonathan P. Gardner}
\affiliation{Astrophysics Science Division, NASA Goddard Space Flight Center, 8800 Greenbelt Rd, Greenbelt, MD 20771, USA}

\author[0000-0003-1530-8713]{Eric Gawiser}
\affiliation{Physics and Astronomy Department, Rutgers, The State University of New Jersey, Piscataway, NJ 08854}

\author[0000-0002-7831-8751]{Mauro Giavalisco}
\affiliation{University of Massachusetts Amherst, 710 North Pleasant Street, Amherst, MA 01003-9305, USA}

\author[0000-0002-5688-0663]{Andrea Grazian}
\affiliation{INAF--Osservatorio Astronomico di Padova, Vicolo dell'Osservatorio 5, I-35122, Padova, Italy}

\author[0000-0001-9440-8872]{Norman A. Grogin}
\affiliation{Space Telescope Science Institute, 3700 San Martin Drive, Baltimore, MD, 21218 USA}

\author[0000-0001-6145-5090]{Nimish P. Hathi}
\affiliation{Space Telescope Science Institute, 3700 San Martin Drive, Baltimore, MD, 21218 USA}

\author[0000-0002-3301-3321]{Michaela Hirschmann}
\affiliation{Institute of Physics, Laboratory of Galaxy Evolution, EPFL, Observatoire de Sauverny, 1290 Versoix, Switzerland}

\author[0000-0002-4884-6756]{Benne W. Holwerda}
\affil{Physics \& Astronomy Department, University of Louisville, 40292 KY, Louisville, USA}

\author[0000-0002-1416-8483]{Marc Huertas-Company}
\affil{Instituto de Astrof\'isica de Canarias, La Laguna, Tenerife, Spain}
\affil{Universidad de la Laguna, La Laguna, Tenerife, Spain}
\affil{Universit\'e Paris-Cit\'e, LERMA - Observatoire de Paris, PSL, Paris, France}

\author[0000-0001-6251-4988]{Taylor A. Hutchison}
\affiliation{Department of Physics and Astronomy, Texas A\&M University, College Station, TX, 77843-4242, USA}
\affiliation{George P.\ and Cynthia Woods Mitchell Institute for Fundamental Physics and Astronomy, Texas A\&M University, College Station, TX, 77843-4242, USA}

\author[0000-0002-1590-0568]{Shardha Jogee}
\affiliation{Department of Astronomy, The University of Texas at Austin, 2515 Speedway, Stop C1400, Austin, TX 78712, USA}

\author[0000-0002-0000-2394]{St\'{e}phanie Juneau}
\affiliation{NSF's National Optical-Infrared Astronomy Research Laboratory, 950 N. Cherry Ave., Tucson, AZ 85719, USA}

\author[0000-0003-1187-4240]{Intae Jung}
\affil{Department of Physics, The Catholic University of America, Washington, DC 20064, USA}
\affil{Astrophysics Science Division, NASA Goddard Space Flight Center, 8800 Greenbelt Rd, Greenbelt, MD 20771, USA}
\affil{Center for Research and Exploration in Space Science and Technology, NASA/GSFC, Greenbelt, MD 20771}

\author[0000-0001-9187-3605]{Jeyhan S. Kartaltepe}
\affiliation{Laboratory for Multiwavelength Astrophysics, School of Physics and Astronomy, Rochester Institute of Technology, 84 Lomb Memorial Drive, Rochester, NY 14623, USA}

\author[0000-0002-5537-8110]{Allison Kirkpatrick}
\affiliation{Department of Physics and Astronomy, University of Kansas, Lawrence, KS 66045, USA}

\author[0000-0002-8360-3880]{Dale D. Kocevski}
\affiliation{Department of Physics and Astronomy, Colby College, Waterville, ME 04901, USA}

\author[0000-0002-6610-2048]{Anton M. Koekemoer}
\affiliation{Space Telescope Science Institute, 3700 San Martin Drive, Baltimore, MD, 21218 USA}

\author[0000-0003-3130-5643]{Jennifer M. Lotz}
\affiliation{Gemini Observatory/NSF's National Optical-Infrared Astronomy Research Laboratory, 950 N. Cherry Ave., Tucson, AZ 85719, USA}

\author[0000-0003-1581-7825]{Ray A. Lucas}
\affiliation{Space Telescope Science Institute, 3700 San Martin Drive, Baltimore, MD, 21218 USA}

\author[0000-0002-6777-6490]{Benjamin Magnelli}
\affiliation{Universit\'e Paris-Saclay, Universit\'e Paris Cit\'e, CEA, CNRS, AIM, 91191, Gif-sur-Yvette, France}

\author[0000-0002-7547-3385]{Jasleen Matharu}
\affiliation{Department of Physics and Astronomy, Texas A\&M University, College Station, TX, 77843-4242, USA}
\affiliation{George P.\ and Cynthia Woods Mitchell Institute for Fundamental Physics and Astronomy, Texas A\&M University, College Station, TX, 77843-4242, USA}

\author[0000-0003-4528-5639]{Pablo G. P\'erez-Gonz\'alez}
\affiliation{Centro de Astrobiolog\'{\i}a (CAB/CSIC-INTA), Ctra. de Ajalvir km 4, Torrej\'on de Ardoz, E-28850, Madrid, Spain}

\author[0000-0003-3382-5941]{Nor Pirzkal}
\affiliation{ESA/AURA, Space Telescope Science Institute, 3700 San Martin Drive, Baltimore, MD, 21218, USA}

\author[0000-0002-9946-4731]{Marc Rafelski}
\affiliation{Space Telescope Science Institute, 3700 San Martin Drive, Baltimore, MD, 21218 USA}
\affiliation{Department of Physics and Astronomy, Johns Hopkins University, Baltimore, MD 21218, USA}
i
\author[0000-0002-8018-3219]{Caitlin Rose}
\affiliation{Laboratory for Multiwavelength Astrophysics, School of Physics and Astronomy, Rochester Institute of Technology, 84 Lomb Memorial Drive, Rochester, NY 14623, USA}

\author[0000-0001-7755-4755]{Lise-Marie Seill\'e}
\affiliation{Aix Marseille Univ, CNRS, CNES, LAM Marseille, France}

\author[0000-0002-6748-6821]{Rachel S.~Somerville}
\affiliation{Center for Computational Astrophysics, Flatiron Institute, 162 5th Avenue, New York, NY 10010, USA}

\author[0000-0002-4772-7878]{Amber N. Straughn}
\affiliation{Astrophysics Science Division, NASA Goddard Space Flight Center, 8800 Greenbelt Rd, Greenbelt, MD 20771, USA}

\author[0000-0002-8224-4505]{Sandro Tacchella}
\affiliation{Kavli Institute for Cosmology, University of Cambridge, Madingley Road, Cambridge, CB3 0HA, UK}
\affiliation{Cavendish Laboratory, University of Cambridge, 19 JJ Thomson Avenue, Cambridge, CB3 0HE, UK}

\author[0000-0002-8163-0172]{Brittany N. Vanderhoof}
\affiliation{Laboratory for Multiwavelength Astrophysics, School of Physics and Astronomy, Rochester Institute of Technology, 84 Lomb Memorial Drive, Rochester, NY 14623, USA}

\author[0000-0001-6065-7483]{Benjamin J. Weiner}
\affiliation{MMT/Steward Observatory, University of Arizona, 933 N. Cherry St, Tucson, AZ 85721, USA }

\author[0000-0003-3735-1931]{Stijn Wuyts}
\affiliation{Department of Physics, University of Bath, Claverton Down, Bath BA2 7AY, UK}

\author[0000-0003-3466-035X]{L. Y. Aaron\ Yung}
\affiliation{Astrophysics Science Division, NASA Goddard Space Flight Center, 8800 Greenbelt Rd, Greenbelt, MD 20771, USA}

\author[0000-0002-0786-7307]{Jorge A. Zavala}
\affiliation{National Astronomical Observatory of Japan, 2-21-1 Osawa, Mitaka, Tokyo 181-8588, Japan}

\begin{abstract}
We present rest-frame optical emission-line flux ratio measurements for five $z>5$ galaxies observed by the \textit{James Webb Space Telescope} Near-Infared Spectrograph (NIRSpec) in the SMACS~0723 Early Release Observations. We add several quality-control and post-processing steps to the NIRSpec pipeline reduction products in order to ensure reliable \textit{relative} flux calibration of emission lines that are closely separated in wavelength, despite the uncertain \textit{absolute} spectrophotometry of the current version of the reductions. Compared to $z\sim3$ galaxies in the literature, the $z>5$ galaxies have similar $\OIII\lambda5008/\Hb$ ratios, {similar $\OIII\lambda4364/\Hg$ ratios, and higher ($\sim$0.5~dex) $\NeIII\lambda3870/\OII\lambda3728$ ratios}. We compare the observations to MAPPINGS~V photoionization models and find that the measured $\NeIII\lambda3870/\OII\lambda3728$, $\OIII\lambda4364/\Hg$, and $\OIII\lambda5008/\Hb$ emission-line ratios are consistent with an interstellar medium that has very high ionization ($\log(Q) \simeq 8-9$, units of cm~s$^{-1}$), low metallicity ($Z/Z_\odot \lesssim 0.2$), and very high pressure ($\log(P/k) \simeq 8-9$, units of cm$^{-3}$). The combination of $\OIII\lambda4364/\Hg$ and $\OIII\lambda(4960+5008)/\Hb$ line ratios indicate very high electron temperatures of $4.1<\log(T_e/{\rm K})<4.4$, further implying metallicities of $Z/Z_\odot \lesssim 0.2$ with the application of low-redshift calibrations for {``$T_e$-based''} metallicities. These observations represent a tantalizing new view of the physical conditions of the interstellar medium in galaxies at cosmic dawn.
\end{abstract}

\section{Introduction}

Emission lines provide a wealth of information about the physical conditions of galaxies. In particular, rest-frame optical lines can reveal the star formation rate \citep{kennicutt12}, nebular dust attenuation \citep{buat02,groves12}, active galactic nucleus (AGN) content \citep{BPT81,VO87}, and the metallicity \citep{lequeux79,tremonti04}, ionization \citep{kewley19review}, and density \citep{dopita00} of the interstellar medium (ISM). Pairs of high-ionization and low-ionization lines that are closely separated in wavelength -- for example, $\NII\lambda6584/\Ha$, $\OIII\lambda5008/\Hb$, $\OIII\lambda4364/\Hg$, and $\NeIII\lambda3870/\OII\lambda3728$ -- are relatively insensitive to dust attenuation and so are especially useful as probes of ISM conditions.

The advent of efficient, multi-object optical and near-infrared spectroscopic surveys has expanded our knowledge of galaxy physical conditions from the local Universe to the peak of cosmic star formation at $z \sim 2$ \citep{madau14}. Galaxies at $1<z<3.5$ have lower metallicity than $z \sim 0$ galaxies of the same stellar mass \citep{henry13,steidel14,maiolino19,sanders21}, as expected from enrichment by star formation. But beyond the metallicity evolution they also have higher ionization \citep{liu08,kewley15,shapley15,strom18,backhaus22,papovich22}. Compared to the current epoch, galaxies at $1<z<2$ have higher AGN content \citep{trump11,juneau14,coil15}, higher-density \HII\ regions \citep{brinchmann08,liu08,davies21}, and more $\alpha$-enrichment from Wolf-Rayet stars and/or massive binaries \citep{masters14,strom17,sanders20}.

About a quarter of the stars in our Universe assemble at $z \gtrsim 2$ \citep{madau14}. Galaxies at these early times are expected to have even more extreme ISM conditions, with lower metallicity and higher ionization observed in their rest-frame UV emission \citep{smit14,stark15,amorin17,stark17,hutchison19}. The observed mid-infrared colors of high-redshift galaxies also suggest contribution from rest-frame optical emission lines with very high equivalent widths \citep{vanderwel11,gonzalez12,smit14,endsley21}, implying high star formation rates and a highly-ionized ISM. But directly measuring rest-frame optical emission lines of galaxies at {$2<z \lesssim 5$ has been enormously challenging} due to high sky background from the ground, and has been {entirely impossible at $z>5$}... until now.

\begin{figure}[t]
\centering
\includegraphics[scale=0.4]{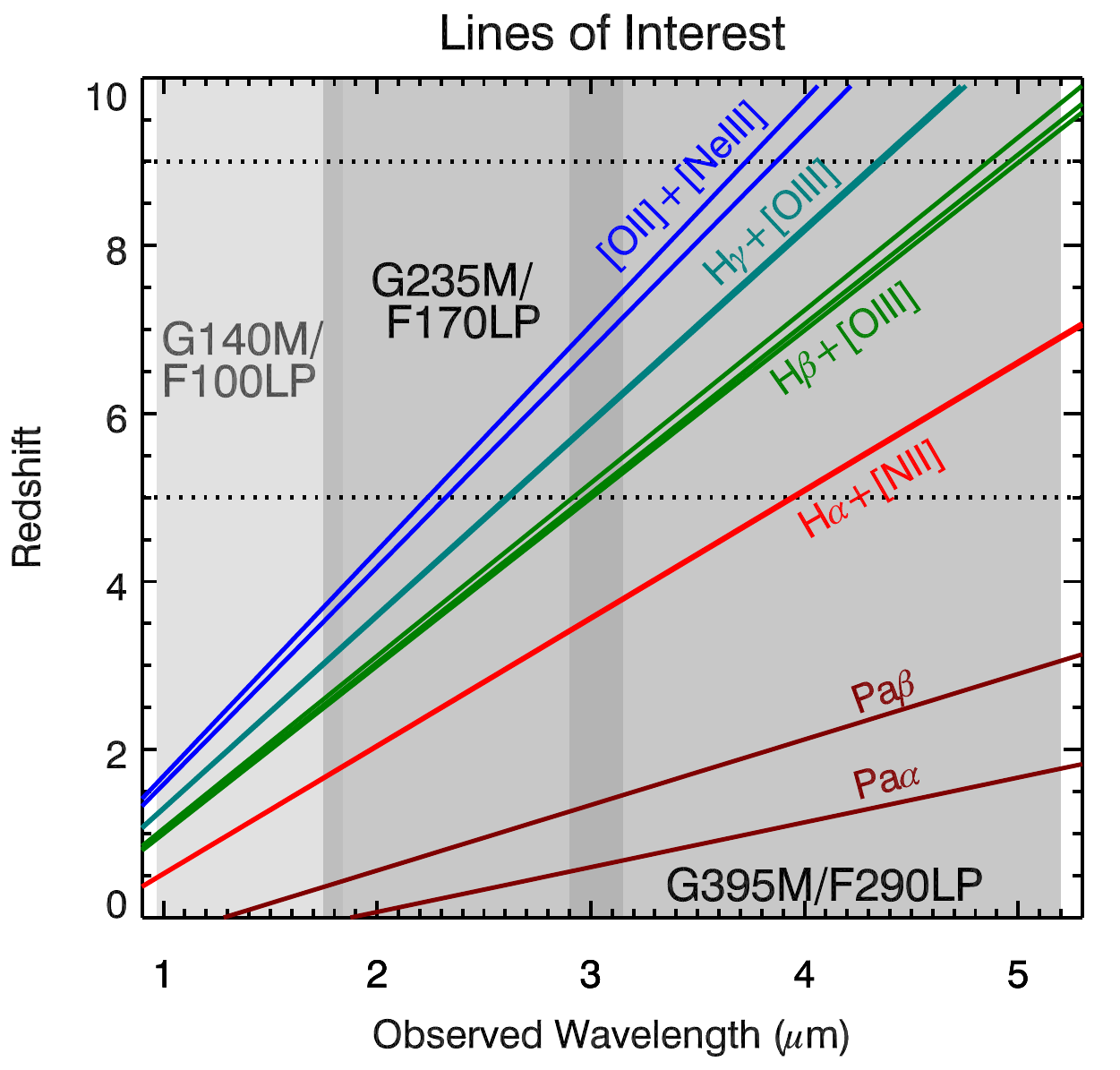}
\caption{An illustration of the emission lines detectable by \jwst/NIRSpec at different redshifts. The gray regions show the observed-frame wavelength range of three medium-resolution grating and filter combinations, with darker gray regions indicating the overlapping wavelength regions covered by two gratings. We do not use the G140M/F100LP grating/filter in this work and so show its wavelength coverage in a lighter shade for illustrative purposes only. In this work we use SMACS~0723 Early Release Observations with G235M/F170LP and G395M/F290LP to study rest-frame optical lines of five galaxies at $5<z<9$.
\label{fig:linecoverage}}
\end{figure}

The launch of the \textit{James Webb Space Telescope} (\jwst, \citealp{jwst}) opens an entirely new window on the high-redshift Universe. \jwst/NIRSpec spectroscopy spans observed-frame 1-5~$\mu$m, enabling detection of rest-frame optical emission lines to $z \lesssim 9$. Figure \ref{fig:linecoverage} highlights the coverage of the \jwst/NIRSpec medium-resolution gratings for various rest-frame optical emission lines as a function of redshift. The advent of \jwst\ observations finally allows a direct comparison of physical conditions over 13~Gyr of cosmic time, using the same set of rest-frame optical emission-line diagnostics from cosmic dawn to the current epoch.

In this paper we investigate the emission-line properties of five $z>5$ galaxies with \jwst/NIRSpec spectroscopy from Early Release Observations of the galaxy cluster SMACS J0723.3–7327 (henceforth SMACS~0723). Section~2 describes the observations and data reduction, which includes some post-processing to ensure reliable relative flux calibration. Section~3 describes our spectral fitting and measurements of emission-line flux ratios. In Section~4 we compare our new $z>5$ line-ratio measurements with previous observations at lower redshift and with theoretical photoionization models, finding that the high-redshift galaxies have very high ionization ($\log(Q/[\mathrm{cm~s}^{-1}]) \sim 8-9$) and low (but nonzero) metallicities ($Z/Z_\odot \sim 0.1$). We summarize the results in Section~5.

\section{Observations}

SMACS~0723 was observed by program \#2736 as part of the \jwst\ Early Release Observations\footnote{https://www.stsci.edu/jwst/science-execution/approved-programs/webb-first-image-observations} {\citep{pontoppidan22}}. In this paper we focus on the NIRSpec observations of the 5 galaxies at $z>5$ that were presented by \citet{carnall22}. All 5 of these galaxies are gravitationally lensed by the foreground cluster: ID 4590 ($z=8.5$) has a magnification factor of 8, while the other 4 galaxies have more modest magnification factors of 1.5-2 (using the parametric model of \citealp{pascale22}). SMACS~0723 was also observed with NIRISS spectroscopy and with NIRCam and MIRI imaging although we do not use those data in this work.

{The \textit{JWST} data used in this paper can be found in MAST: \dataset[10.17909/67ft-nb86]{http://dx.doi.org/10.17909/67ft-nb86}.}


\subsection{NIRSpec Observational Setup}

The details of the NIRSpec instrument {and the microshutter array (MSA)} are described by \citet{jakobsen22} {and \citet{ferruit22}, respectively.}

SMACS~0723 was observed with the G235M/F170LP (1.75-3.15$\mu$m) and G395M/F290LP (2.9-5.2$\mu$m) grating/filter pairs, each of which has spectral resolution of $R \simeq 1000$. Each grating was observed with two NIRSpec visits, with each visit using a three-nod pattern and two integrations of 20 groups (2918~s) per nod. The coadded spectra from each visit (combining from the three nods) have a total exposure time of 8754~s in each grating. Targets for the MSA configuration were selected using the NIRCam imaging in the field, especially prioritizing targets with photometric redshifts of $z>6$. Each target was observed using a ``slitlet'' aperture of three microshutters and the design also included empty shutters for background subtraction.

\subsection{Data Reduction and Quality Checks}

{We perform a complete reduction from Level~0 raw uncalibrated data (``\_uncal.fits'' files)} available on the Mikulski Archive for Space Telescopes server (MAST)\footnote{https://mast.stsci.edu/}, {processed using version~1.8.2 of the \jwst\ Science Calibration Pipeline with the ``jwst-1015.pmap'' calibration context. This reduction includes updates to the instrument models, gain response, flats and detector-level calibration files released in the months following the observations of the ERO programs.}
The reduced two-dimensional (2D) spectra (``s2d'') have a rectified trace with a flat slope, drizzled from the significantly curved (by 14-24 pixels over the spectral range) trace observed on the detector. The pipeline-reduced one-dimensional (1D) spectra (``x1d'') are extracted from the 2D spectra using an ``extended'' aperture with a width of 8~pixels. We instead extract spectra using a narrower ``point-source'' aperture, as described below.

\begin{figure}[t]
\centering
\epsscale{1.1}
\plotone{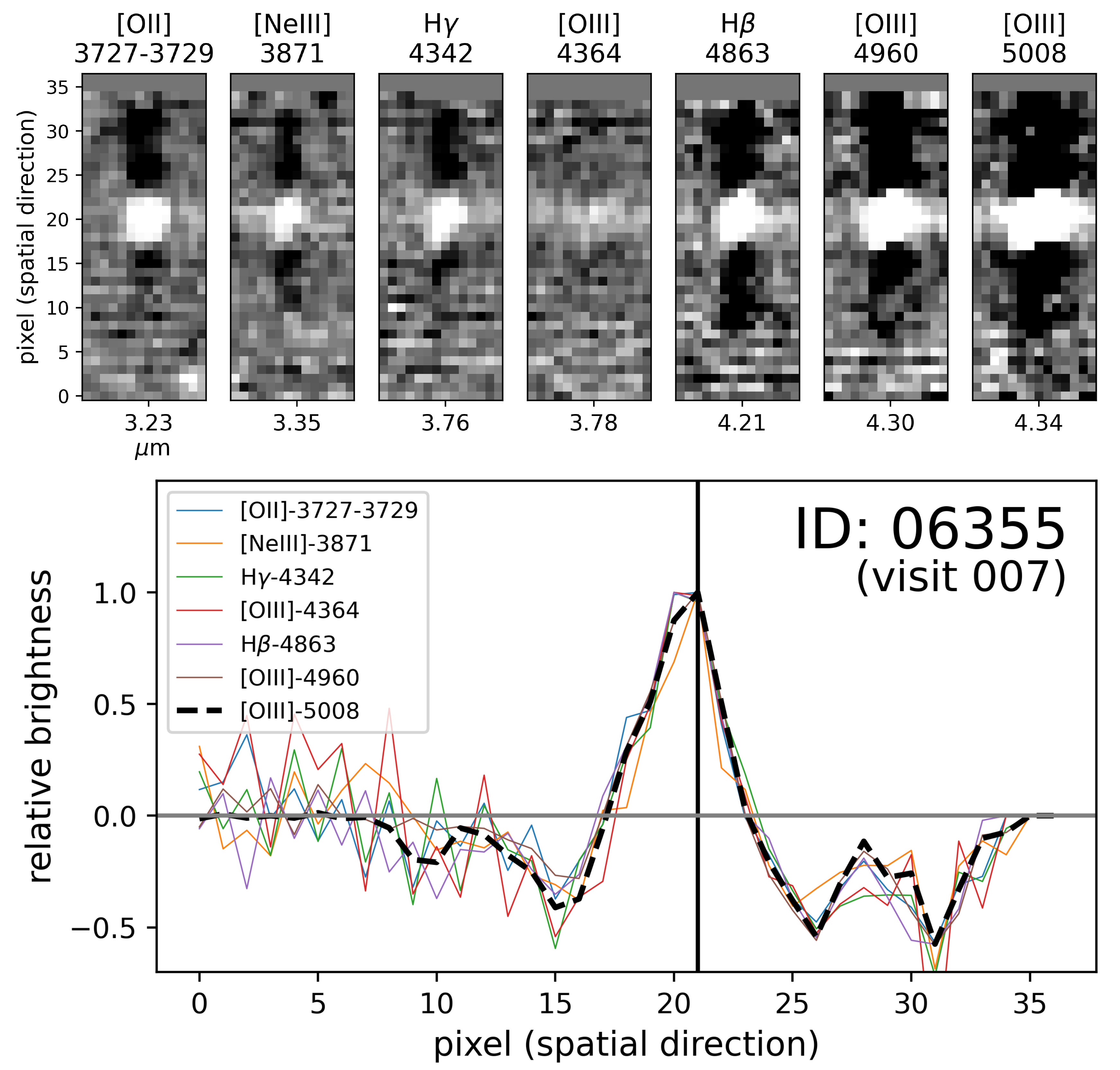}
\caption{{The 2D profiles} of the rest-frame optical emission lines of interest for the first visit of source 6355. The pixel scale is $0\farcs1$/pixel in the cross-dispersion (vertical) direction and 17~\AA/pixel in the wavelength (horizontal) direction. The darker regions are individual nod positions and the brighter regions are from the coadd of the three nods. The emission lines are well-detected and have a consistent spectral trace over a broad range of wavelength in the reduced 2D spectra. The complete figure set (10 images) of all 10 sets of emission-line profiles (2 visits each for 5 sources) is available in the online journal.
\label{fig:lineprofiles}}
\end{figure}

We confirmed that the reduced 1D spectra have excellent wavelength calibration, with differences of $\Delta z / z \lesssim 10^{-4}$ in best-fit line centers for different emission lines across the observed spectral range. We also confirmed that the reduced 2D spectra have a flat trace, with consistent spatial profiles of emission lines over the spectral range of each grating. The 2D line profiles for the galaxies are shown in Figure \ref{fig:lineprofiles}.

{The current (version~1.8.2) data reduction pipeline uses a flux calibration that
relies on knowledge of the instrument before launch. The pre-launch instrument throughput is known to differ from the post-launch performance (see Figure~20 of \citealt{jwstperformance}). We also find that synthetic photometry from the spectra have a median difference of $\sim$30\% from the NIRCam photometry in the F200W, F356W, and F444W filters (and a smaller median difference of 14\% in the F277W filter).
For these reasons we avoid analysis and interpretation that require \textit{absolute} spectrophotometric calibration, like using individual emission line fluxes or widely separated line ratios (e.g., \OIII/\OII). We confirm below that the spectra have reliable \textit{relative} spectrophotometry for pairs of emission lines that are closely separated in wavelength.}

The {default} 8-pixel ``extended'' extraction width used by the current (version~1.8.2) pipeline is generally too large for the compact high-redshift sources that are the focus of this paper: Figure \ref{fig:lineprofiles} shows that all of our sources have their spectral trace confined to 4-5~pixels. We also found that the ``extended'' extraction apertures were often not well-centered on the target, such that the 1D spectrum included a significant amount of the ``negative-nod'' flux above and below the coadded 2D spectrum. The 1D spectra produced by these wide extraction apertures often include emission from serendipitous sources and detector artifacts that lie outside the trace of the main source. The wide-extraction spectra generally have inconsistent emission-line flux measurements between the two visits that differ by up to a factor of $\sim$2, with even worse (factor of $\sim$10) differences in the continuum emission. In addition, the wide-extraction spectra often have unphysical Balmer line ratios: e.g., $\Hb/\Hg \sim 1$ compared to atomic calculations of $\Hb/\Hg \simeq 2.1$ for a broad range of temperature and density \citep{osterbrock89}. {Other work on this same dataset \citep{schaerer22,curti22,taylor22,brinchmann22}} noted some of the same issues with the flux calibration of the pipeline-reduced data and took independent approaches to mitigating them.

\begin{figure*}[t]
\centering
\epsscale{1.0}
\plotone{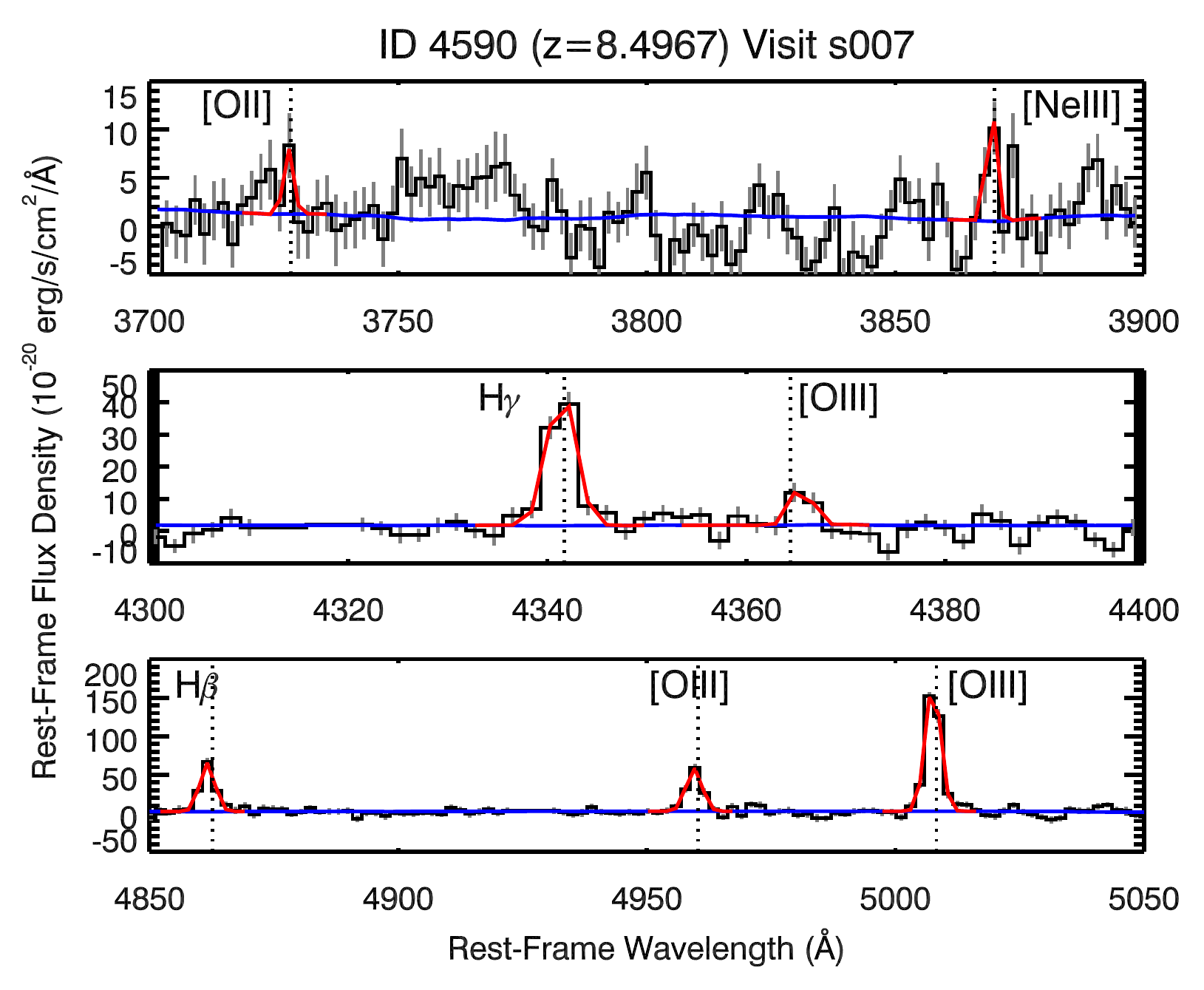}
\caption{The spectrum and emission-line fits for the first visit of source 4590, in the rest-frame wavelength regions that include \OII\ and \NeIII, \Hg\ and $\OIII\lambda4364$, \Hb\ and $\OIII\lambda4960,5008$. Flux is shown by the black histogram and flux uncertainty by the gray error bars. The blue line indicates the continuum and the red lines indicate the best-fit Gaussian functions to each emission line. The complete figure set (10 images) of all 10 spectra and emission-line fits (2 visits each for 5 sources) is available in the online journal.
\label{fig:linefits}}
\end{figure*}

\begin{deluxetable*}{lrc|ccccccc}[t]
\tablenum{1}
\tablecolumns{10}
\tablecaption{Target ID, Redshift, and Emission-Line {SNR} \label{tbl:sample}}
\tablehead{
\colhead{ID} & \colhead{Redshift} & \colhead{Visit} & \colhead{\OII} & \colhead{\NeIII} & \colhead{\Hg} & \colhead{\OIII$\lambda$4364} & \colhead{\Hb} &  \colhead{$\OIII\lambda4960$} & \colhead{$\OIII\lambda5008$}
}
\startdata
 4590 & 8.4957 & 7 &  2.1 &  3.9 & 11.9 &  3.8 & 14.1 & 11.2 & 24.2 \\
 4590 & 8.4957 & 8 &  2.6 &  5.6 &  8.8 &  3.5 & 14.9 &  *** & 29.8 \\
 5144 & 6.3792 & 7 &  5.2 &  8.0 & 11.0 &  2.8 & 20.5 & 31.9 & 62.9 \\
 5144 & 6.3792 & 8 &  5.3 &  7.3 & 10.7 &  5.1 & 18.4 & 32.9 & 59.3 \\
 6355 & 7.6651 & 7 & 17.2 & 11.1 & 11.5 &  3.1 & 19.0 & 36.0 & 70.6 \\
 6355 & 7.6651 & 8 & 16.2 & 10.8 &  8.8 &  2.1 & 18.2 & 33.7 & 66.2 \\
 8140 & 5.2753 & 7 &  9.0 &  2.4 &  0.8 &  *** &  6.1 & 11.5 & 21.0 \\
 8140 & 5.2753 & 8 & 10.4 &  4.0 &  1.0 &  1.3 &  4.3 &  9.5 & 19.4 \\
10612 & 7.6597 & 7 &  4.2 &  9.5 &  9.5 &  5.1 & 17.5 & 29.9 & 57.4 \\
10612 & 7.6597 & 8 &  *** &  9.1 &  9.4 &  4.2 & 15.8 & 27.6 & 52.5
\enddata
\caption{{Emission lines are measured independently for each of the 2 NIRSpec visits for each source. Asterisks (***) indicate a problem in the spectrum (a detector artifact or other emission beyond the main spectral trace) that prevents a measurement of the emission line.
}}
\end{deluxetable*}

\begin{deluxetable*}{lr|cccc}[t]
\tablenum{2}
\tablecolumns{6}
\tablecaption{{Emission-Line Ratios \label{tbl:ratios}}}
\tablehead{
\colhead{ID} & \colhead{Redshift} & 
\colhead{\NeIII/\OII} & \colhead{$\OIII\lambda4364/\Hg$} &
\colhead{$\OIII\lambda5008/\Hb$} & 
\colhead{$\OIII\lambda5008/\OIII\lambda4960$}
}
\startdata
 4590 & 8.4957 & $1.82 \pm 0.63$ & $0.28 \pm 0.06$ & $3.05 \pm 0.17$ & $2.85 \pm 0.28$ \\
 5144 & 6.3792 & $1.32 \pm 0.22$ & $0.27 \pm 0.05$ & $6.45 \pm 0.25$ & $3.04 \pm 0.08$ \\
 6355 & 7.6651 & $0.48 \pm 0.04$ & $0.21 \pm 0.06$ & $8.23 \pm 0.32$ & $3.10 \pm 0.07$ \\
 8140 & 5.2753 & $0.39 \pm 0.09$ &     ***         & $6.82 \pm 0.98$ & $2.85 \pm 0.22$ \\
10612 & 7.6597 & $1.84 \pm 0.48$ & $0.37 \pm 0.06$ & $6.97 \pm 0.31$ & $2.99 \pm 0.08$
\enddata
\caption{{Line ratios are measured from the average of the two visits (excepting the cases where a line cannot be measured in one visit). Error bars indicate 1$\sigma$ uncertainties. The $\OIII\lambda4364/\Hg$ ratio cannot be measured for ID~8140 and is marked by asterisks (***).}}
\end{deluxetable*}

\vspace{-1.75cm}
Rather than using the wide-extraction 1D spectra from the pipeline, we produce new 1D spectra from a narrower ``point-source'' extraction width {individually optimized for each source (typically $\sim$4 pixels wide)}. This required significant customization of the (version~1.8.2) NIRSpec reduction pipeline in order to accurately align the extraction window with the position of the source. These narrow-extraction 1D spectra represent a dramatic improvement over the wide-extraction versions: they avoid much of the contamination from serendipitous sources and detector artifacts and have emission-line fluxes that are 2-4$\times$ larger due to avoiding the negative-nod emission present in the ``extended'' apertures. Most importantly, they have consistent emission-flux measurements between visits (with one exception noted in Section~3).
{We also visually inspect the 1D and 2D spectra and mask obvious defects in the spectra, generally caused by chip gaps or bad pixels on the CCD.}

The flux uncertainties of our reduced 1D spectra appear to be underestimated by a factor of $\sim$2 (and by a factor of $\sim$1.3 in the wide-extraction spectra), as measured from a comparison of the normalized median absolute deviation (NMAD) of the flux with the median of the flux uncertainty for each source, calculated in wavelength regions without emission lines and avoiding chip gaps and bad pixels. We increase the flux uncertainty of the spectra using the ratio of the NMAD of the flux to the median flux uncertainty, i.e. an error rescaling factor of $\mathrm{NMAD}(f) / \mathrm{median}(\sigma_f)$. We note that this error rescaling may still remain an underestimate of the true noise if the pixels of the spectrum are correlated.

Our post-processing improvements in flux calibration and 1D extraction represent a significant improvement over the {(version~1.8.2)} pipeline-reduced 1D spectra. However, our flux calibration additionally relies on pre-launch knowledge of the instrument that differs from the measured post-launch performance (Figure~20 of \citealt{jwstperformance}). In addition, the wavelength-dependent spatial resolution of NIRSpec will cause wavelength-dependent effects from aperture losses when using a fixed-width 2D spectral extraction. Despite these potential problems in the \textit{absolute} flux calibration, we find that the \textit{relative} flux calibration is consistent for pairs of emission lines that are near one another in wavelength. The line ratios of near-pair lines are also consistent between visits, as discussed in the next Section. Thus we are confident in using ratios of emission lines that are closely separated in wavelength, but we caution against the use of emission-line fluxes and equivalent widths, and against the direct comparison of lines that are widely separated in wavelength (like $\OIII\lambda4364/\OIII\lambda5008$).

\section{Emission-Line Flux Ratio Measurements}

We fit for the following emission lines in each spectrum (noted by vacuum wavelengths in Angstroms):
\begin{itemize}
    \setlength\itemsep{0em}
    \item $\OII\lambda3728.48$ (the 3727+3729 doublet is blended in the $R \simeq 1000$ medium-resolution NIRSpec grating)
    \item $\NeIII\lambda3870.86$
    \item $\Hg\lambda4341.69$
    \item $\OIII\lambda4364.44$
    \item $\Hb\lambda4862.72$
    \item $\OIII\lambda4960.30$
    \item $\OIII\lambda5008.24$
\end{itemize}

We find the best-fit Gaussian function {(and associated uncertainties)} for each emission line using a Levenberg-Marquardt least-squares method implemented by the \texttt{mpfit} IDL code\footnote{https://pages.physics.wisc.edu/$\sim$craigm/idl/fitting.html}. We subtract a continuum that is determined by smoothing (by a boxcar of 100 pixels) and interpolating the flux from all regions that are $<$5$\sigma$ above the median flux of the spectrum (i.e., over line-free regions). We fit the spectra from each of the two visits independently. Examples of the emission-line fits are shown in Figure \ref{fig:linefits}. In a few cases, a {line flux cannot be measured} due to {contaminating emission that extends beyond the main spectral trace (likely from a detector artifact or serendipitous source)}:
\begin{itemize}
    \setlength\itemsep{0em}
    \item ID 4590, second visit: $\OIII\lambda4960$
    \item ID 8140, first visit: $\OIII\lambda4364$
    \item ID 10612, second visit: $\OII\lambda3728$
\end{itemize}

{Table \ref{tbl:sample} presents the source IDs, spectroscopic redshifts, and signal-to-noise (SNR) of each emission-line measurement for each visit of the observations. Table \ref{tbl:ratios} presents the measured line ratios, generally computed from the average line measurements of the two visits. The line ratio is measured from only one visit if a line of the ratio cannot be measured in the other visit, i.e. $\OIII\lambda5008/\OIII\lambda4960$ for ID~4590 and $\NeIII/\OII$ for ID~10612. Neither $\OIII\lambda4364$ nor \Hg\ are robustly ($>$3$\sigma$) detected in ID~8140 and so the $\OIII\lambda4364/\Hg$ line ratio is unconstrained for this galaxy.}

We determined the spectroscopic redshift for each source using the best-fit line center for $\OIII\lambda5008$, which was the brightest emission line in each spectrum. As noted in Section~2.2, the reduced NIRSpec 1D spectra have excellent wavelength calibration and we found differences of only $\Delta z / z \lesssim 10^{-4}$ when measuring the redshift from the line centers of other emission lines.

Most of the emission lines are measured from the G395M spectrum, with bluer lines measured in the G235M spectrum for the lower-redshift sources. Because {comparison with NIRCam photometry of the same galaxies indicates that} the absolute flux calibration is suspect by a factor of $\sim$30\%, we use only ratios of near-pair emission lines rather than individual emission-line fluxes. In cases where an emission line is measured in the wavelength range $2.9<\lambda\,(\mu\mathrm{m})<3.2$ that is covered by both gratings, we take care to measure {both lines in a given ratio from the same grating}, given the differences in line flux measured from each grating (noted in Section~2.2).

The measured line strengths are generally consistent within their uncertainties between the two visits. {ID~4590 is an exception, with a a factor of $\sim$2 difference in the measured line fluxes between the two visits. This is caused by a shutter that failed to open in one of the nod positions of the first (007) visit, as identified by \citet{curti22}}. {Excluding ID~4590, the ratio of emission-line strengths measured in each visit is $0.99 \pm 0.14$ (mean and standard deviation of the sample) for lines that are $>$3$\sigma$ detected.} Note that despite the difference in line fluxes, the emission-line \textit{ratios} of ID~4590 are consistent between the two visits. Table \ref{tbl:ratios} also demonstrates that the measured {$\OIII\lambda5008/\OIII\lambda4960 = 2.98 \pm 0.12$ (mean and standard deviation of the sample)}, matching the atomic physics calculation \citep{storey00} and establishing the reliability of the relative flux calibration for near-wavelength line pairs.


\begin{figure*}[t]
\centering
\epsscale{0.9}
\plotone{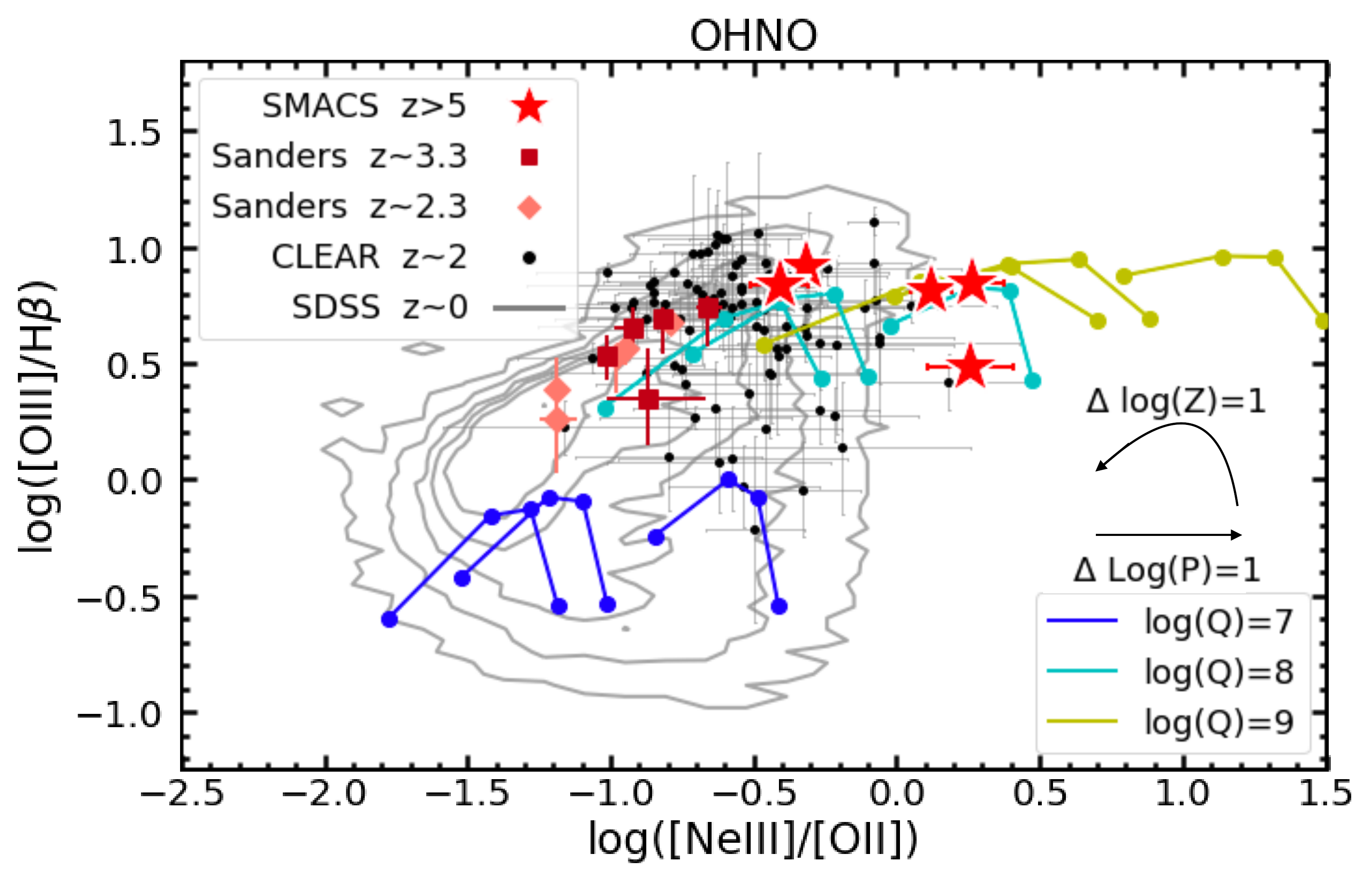}
\caption{The ``OHNO'' diagram of $\OIII\lambda5008/\Hb$ vs.\ $\NeIII\lambda3870/\OII\lambda3728$. As indicated in the legend, the new high-$z$ galaxy measurements are shown by red stars, $z \sim 2$ galaxies from CLEAR in black \citep{backhaus22}, and $z \sim 0$ galaxies from SDSS in gray contours \citep{sdss}.
MAPPINGS models are shown by the colored curves, with different curves for different ionization ($\log(Q)=[7,8,9]$ increasing left to right),  metallicity along each curve ($Z/Z_\odot=[1,0.4,0.2,0.05]$ decreasing left to right), and curves shown for each of three pressures ($\log{P/k}=[7,8,9]$). The OHNO line ratios of the $z>5$ galaxies indicate higher ionization, lower metallicity, and higher pressure than the $z \sim 2$ (and $z \sim 0$) galaxies, and are broadly consistent with an ISM of $\log(Q) \simeq 8-9$, $\log(P/k) \simeq 8-9$, {and/or} $Z/Z_\odot \lesssim 0.2$.
\label{fig:ohno}}
\end{figure*}

\begin{figure*}[t]
\centering
\epsscale{0.9}
\plotone{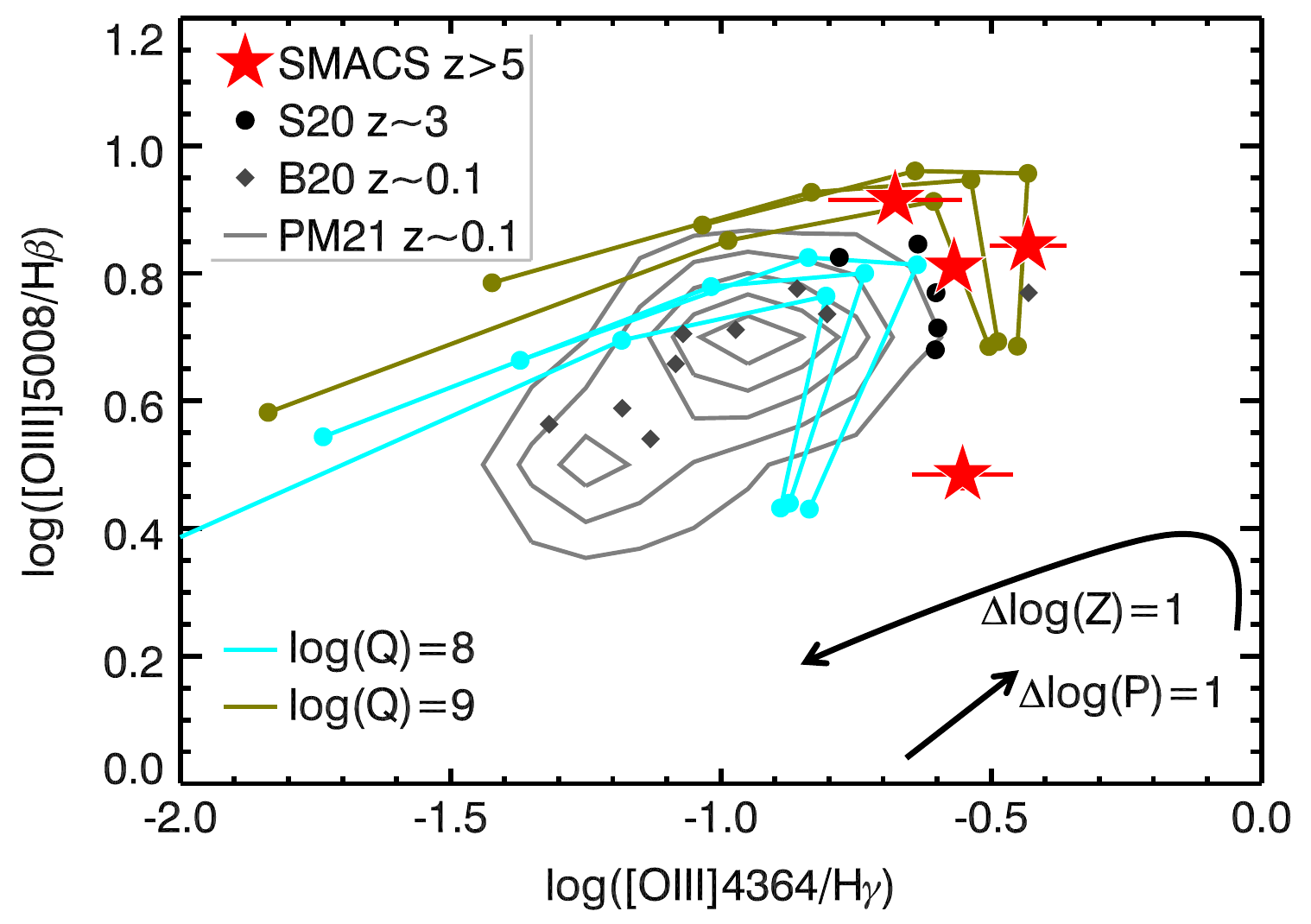}
\caption{The ratio of $\OIII\lambda5008/\Hb$ vs.\ $\OIII\lambda4364/\Hg$. Red stars indicate the new high-$z$ galaxy line ratios. Lower-redshift comparison samples are also shown in the figure: 5 $z \sim 3$ galaxies from \citet{sanders20} (black points), 9 $z \sim 0.35$ ``green-pea'' galaxies from \citet{brunker20} (dark gray diamonds), and $\sim$1800 extreme equivalent-width SDSS galaxies from \citet{perezmontero21} (gray contours). Colored curves indicate theoretical MAPPINGS models for different ionization, metallicity, and pressure. The $\OIII\lambda5008/\Hb$ and $\OIII\lambda4364/\Hg$ ratios are not very sensitive to pressure and so the three $\log{P/k}=[7,8,9]$ models lie very close together. As in Figure \ref{fig:ohno}, the $z>5$ galaxies have $\OIII\lambda5008/\Hb$ ratios that are similar to the lower-redshift comparison sample. The high-redshift $\OIII\lambda4364/\Hg$ ratios are higher than most (but not all) of the $z \sim 0$ galaxies but only slightly ($\lesssim$0.5~dex) larger than the $z \sim 3$ sample. The $z>5$ line ratios are well-described by MAPPINGS models for an ISM of very high ionization ($\log(Q) \simeq 8-9$) and low metallicity ($Z/Z_\odot \lesssim 0.2$).
\label{fig:o3comp}}
\end{figure*}

\section{Line-Ratio Diagnostics}

We infer galaxy properties from emission-line pairs that are closely separated in wavelength: namely $\OIII\lambda5008/\Hb$, $\OIII\lambda4364/\Hg$, and $\NeIII\lambda3870/\OII\lambda3728$. The use of ratios of emission lines that are closely separated in wavelength avoids the issues with the absolute flux calibration described in Section~2.2, and is also largely insensitive to dust attenuation.

In each subsection below, we compare the observations with model spectra from \citet{kewley19}, produced using the MAPPINGS~V photoionization code \citep{mappings}. These models use input stellar ionizing spectra from Starburst99 \citep{starburst99}, which use a \citet{salpeter55} initial mass function and include stellar mass loss. MAPPINGS~V uses atomic data from the CHIANTI~8 database \citep{chianti1,chianti} and applies photoionization, recombination, excitation, and dust depletion of model \HII\ regions in a plane-parallel geometry for the ionizing spectra. We use the ``Pressure Models'' of \citet{kewley19} for a grid of pressure $\log(P/k)$, ionization\footnote{We quantify ionization using $Q=\frac{L_{H^0}}{4\pi R^2 n_H}$, noting that many papers also use $U = Q/c$ (or $\log U = \log [Q/(cm~s^{-1})] - 10.48$) for ionization.} $\log(Q)$, and metallicity $Z/Z_\odot$:
\begin{itemize}
    \setlength\itemsep{0em}
    \item Pressure $\log(P/k) = [7,8,9]$, units of cm$^{-3}$
    \item Ionization $\log(Q) = [7,8,9]$, units of cm~s$^{-1}$
    \item Metallicity $Z/Z_\odot = [0.05,0.2,0.4,1.0]$
\end{itemize}

{The MAPPINGS~V models are characterized in terms of the total metallicity $Z$ with respect to solar, but relative abundances of each element are not simply scaled from the solar abundances. At low metallicities the models use $\alpha$-enhanced abundances as described in \citet{nicholls17}: for example, the relative [O/Fe] abundance is 0.5~dex higher than solar for $Z/Z_\odot<0.1$. The alpha-enhancement at low metallicity in the MAPPINGS~V models is motivated by studies of stellar abundances \citep[e.g.,][]{amarsi19} and is also similar to the alpha-enhancement observed in nebular emission from both low-metallicity galaxies at $z \gtrsim 2$ \citep[e.g.,][]{steidel16,topping20,cullen21} and from a more detailed study of relative abundances in our $z>5$ galaxies \citep{arellanocordova22}}

The ionizing spectra of low-metallicity stars are not well-constrained by observations and at $Z/Z_\odot = 0.05$ the spectrum is essentially extrapolated from the Starburst99 inputs. That means the model spectra are most uncertain at the lowest metallicities, although they generally appear to be smooth continuations of the better-constrained models with higher metallicity. The MAPPINGS~V models also use a single plane-parallel geometry for the ionization and may not effectively model \HII\ regions with multi-phase pressure and ionization and/or more complex geometries \citep{xiao18,kewley19review}.

\subsection{OHNO: $\OIII/\Hb$ and $\NeIII/\OII$}

Figure~\ref{fig:ohno} presents the ``OHNO'' line-ratio diagnostic of $\OIII/\Hb$ vs.\ $\NeIII/\OII$, with the line-ratio measurements of the $z>5$ galaxies shown by large red stars. We use the samples of \citet{backhaus22} as a low-redshift comparison, with line ratios for $z \sim 2$ galaxies in the CLEAR survey (Simons et al.\ in prep)
measured from \hst/WFC3 grism spectroscopy. {A sample of $\sim$28,000 $z \sim 0$ galaxies with detected OHNO emission lines from the Sloan Digital Sky Survey (SDSS, \citealp{sdss}) is shown by gray contours. Figure~\ref{fig:ohno} also shows stacked line-ratio measurements from MOSDEF observations \citep{sanders21} as pink and maroon points.}
{The high- and low-redshift samples have different line-luminosity selection limits and so inter-comparison is nontrivial. The lack of robust absolute flux calibration means that we cannot construct samples of low-redshift galaxies that are matched to our $z>5$ galaxies \citep[following, e.g.,][]{juneau14,backhaus22}. Instead, we generally focus our discussion below on comparing the $z>5$ galaxies to the most extreme (highest ionization and lowest metallicity) galaxies present in the low-redshift samples.}

Compared to the lower-redshift comparison samples, the $z>5$ galaxies in SMACS~0723 have similar $\OIII/\Hb$ ratios but have $\NeIII/\OII$ ratios that are higher by $\sim$0.5~dex. The redshift evolution of $\NeIII/\OII$ {at fixed $\OIII/\Hb$} appears to be broadly consistent from $z \sim 0$ to $z \sim 2$ {to $z \sim 3$} to $z>5$, with {$z \sim 3$ MOSDEF and} $z \sim 2$ CLEAR $\NeIII/\OII$ ratios that are higher than the ``evolution-matched'' $z \sim 0$ sample and $z>5$ $\NeIII/\OII$ ratios that are even higher than the $z \sim 2$ {and $z \sim 3$)} ratios. The redshift evolution of $\NeIII/\OII$ from $z \sim 0$ to $z \sim 2$ {and $z \sim 3$} has been discussed in previous work {\citep[e.g.][]{zeimann15,strom17,kewley19review,jeong20,sanders21,backhaus22} and requires a harder ionizing spectrum, likely caused by some combination of massive $\alpha$-enhanced low-metallicity stars, higher-density (and higher-pressure) \HII\ regions, and increased AGN content at higher redshift.}

Here we demonstrate the same trend of increasing $\NeIII/\OII$ with redshift in galaxies at $z>5$.
{The redshift evolution cannot be explained by the observed anticorrelation of $\NeIII/\OII$ with stellar mass \citep[e.g.,][]{sanders21,backhaus22} since the lowest-mass SDSS, CLEAR, and MOSDEF galaxies have lower (by $\sim$0.5~dex) $\NeIII/\OII$ than the low-mass $z>5$ galaxies.}
We note that there is not an obvious trend of $\NeIII/\OII$ with redshift among the $z>5$ galaxies: for example a $z=7.7$ (ID 6355) galaxy has the lowest $\NeIII/\OII$ and a $z=6.4$ (ID 5144) galaxy has the highest $\NeIII/\OII$, with ratios that are $>$3$\sigma$ inconsistent given their observational uncertainties. This likely indicates a diversity of $\NeIII/\OII$ ratios in individual $z>5$ galaxies, perhaps associated with the diversity of stellar mass, star formation rate, {abundances,} and/or age among this sample. 

Figure~\ref{fig:ohno} additionally compares the observed $z>5$ line ratios with MAPPINGS~V theoretical models, as described above. In general the $z>5$ line ratios are well-described by MAPPINGS~V models with an ISM that is highly ionized ($\log(Q) \simeq 8-9$), high pressure ($\log(P/k) \simeq 8-9$), {and/or} low metallicity {($Z/Z_\odot \lesssim 0.2$)}, {with some degeneracy between the three quantities to best describe the observations}.

\subsection{$\OIII\lambda5008/\Hb$ and $\OIII\lambda4364/\Hg$}

Figure~\ref{fig:o3comp} presents a comparison of the measured $\OIII\lambda5008/\Hb$ and $\OIII\lambda4364/\Hg$ line ratios of the four $z>6$ galaxies in our sample (shown again as large red stars). {ID~8140 ($z=5.3$) is not shown in this Figure because its $\OIII\lambda4364$ and \Hg\ lines are only marginally ($<$2$\sigma$) detected.} Also shown in Figure~\ref{fig:o3comp} are line ratios of 5 $z \sim 3$ galaxies measured by \citet{sanders20}. We also compare to a low-redshift ($0.3<z<0.4$) sample of ``green pea'' galaxies from \citet{brunker20} that are selected based on their high-EW emission lines and are potential analogs to high-redshift / high-ionization starburst galaxies (see also, e.g., \citealp{henry15,flury22}). In addition, the gray contours in Figure \ref{fig:o3comp} show the line-ratio distribution of $\sim$1800 extreme equivalent-width galaxies at $z \sim 0.1$ identified from SDSS observations by \citet{perezmontero21}.

As similarly noted in Section~4.1, the $z>5$ galaxies have similar (high) $\OIII\lambda5008/\Hb$ ratios to the $z \sim 3$ and $z \sim 0$ comparison samples. The $z>5$ galaxies have $\OIII\lambda4364/\Hg$ line ratios that are $\sim$0.5~dex higher than the general distribution of the low-redshift galaxies, although one of the green-pea galaxies has ratios as extreme as the $z>5$ galaxies. {The similarity of these high-redshift galaxies to low-redshift (high-ionization) green-pea galaxies was also noted in other work \citep{katz23,rhoads22}.} The $\OIII\lambda4364/\Hg$ ratios at $z>5$ are {similar to, or perhaps $\sim$0.1-0.2~dex higher, than those of galaxies} at $z \sim 3$. In the next subsection we describe how the high-redshift $\OIII\lambda4364/\Hg$ and $\OIII\lambda5008/\Hb$ line ratios are indicative of very high electron temperatures and low metallicities.

Figure~\ref{fig:o3comp} also compares the observations to the MAPPINGS~V model line ratios (shown by the colored curves of ionization, metallicity, and pressure). {The $\log(Q)=7$ curve lies outside the range of the Figure (to the lower left) and is not shown.} The $\OIII\lambda5008/\Hb$ and $\OIII\lambda4364/\Hg$ line ratios are largely insensitive to ISM pressure and so the three curves of different pressure ($\log(P/k)=[7,8,9]$) have very similar line ratios. As in the OHNO diagram in Figure \ref{fig:ohno}, the observed line ratios in Figure \ref{fig:o3comp} are consistent with the MAPPINGS models for a highly ionized and low-metallicity ISM, with $\log(Q) \simeq 8-9$ and $Z/Z_\odot \lesssim 0.2$.

\subsection{Electron Temperature and Metallicity}

The ratio of the $\OIII\lambda4364$ and the $\OIII\lambda4960+5008$ doublet can be used to measure the electron temperature of the ISM. These lines are all collisionally excited and the $\OIII\lambda4364$ line de-excites from a higher energy orbital, such that higher $\OIII\lambda4364$ emission relative to $\OIII\lambda4960+5008$ implies higher energy electrons responsible for the collisional excitation. {The electron temperature can be used with the $\OIII\lambda4960+5008$, $\OII\lambda3728$, and Balmer lines for a ``direct'' metallicity estimate \citep[e.g.,][]{izotov06,perezmontero17,nicholls20}, although this requires good flux calibration between the widely separated \OII\ and \OIII\ lines. In this work we use} empirical correlations that have been found between electron temperature and the ``direct'' metallicity {\citep{amorin15,perezmontero21} to measure ``$T_e$-based'' metallicities.}

We cannot directly compare our measured $\OIII\lambda4364$ and $\OIII\lambda4960+5008$ line fluxes because of the uncertain absolute flux calibration of the NIRSpec spectra (see Section~2.2 for details). Instead, we rely on the reliable \textit{relative} flux calibration and use the measured ratios of $\OIII\lambda4364/\Hg$ and $\OIII\lambda(4960+5008)/\Hb$, along with the intrinsic {(relatively insensitive to temperature)} Balmer ratio $\Hb/\Hg = 2.1$ \citep{osterbrock89}. In other words, we measure the \OIII\ ratio as follows, abbreviating the \OIII\ lines by their wavelengths:
\begin{equation}
    \frac{\lambda4364}{\lambda(4960+5008)} = \frac{\lambda4364}{\Hg} \left(\frac{\lambda4960+5008}{\Hb}\right)^{-1} \times (2.1)^{-1}
\end{equation}
Note that this method also implicitly corrects the $\OIII\lambda4364/\OIII\lambda4960+5008$ ratio for dust attenuation.

We use Equation 4 of \citet{nicholls20} to estimate $T_e$ and Equation 1 of \citet{perezmontero21} to estimate metallicity from the electron temperature\footnote{\citet{nicholls20} and \citet{perezmontero21} use different atomic datasets, \citet{lennon94} and \citet{storey14}, respectively, but the two datasets are very similar, with only minor differences that are much smaller than the observational uncertainties for the lines used in this work.}.
{We use a $T_e$-based metallicity because it can be calculated solely from near-pair line ratios (Equation 1), as opposed to the ``direct'' metallicity that additionally requires use of the \OII\ line and, by extension, robust absolute flux calibration. Uncertainties are calculated for both $T_e$ and metallicity using Monte Carlo resampling of the line ratios, including the calibration uncertainties for $T_e$-based metallicity reported in Equation 1 of \citet{perezmontero21}.}

The inferred electron temperatures and {$T_e$-based} metallicities, {and their 1$\sigma$ uncertainties}, are shown in Table \ref{tbl:teffz}. {Electron temperature and metallicity cannot be calculated for ID~8140 ($z=5.3$) because its $\OIII\lambda4364/\Hg$ line ratio is unconstrained.} Our electron temperature measurements agree very well with {other studies of these galaxies \citep{schaerer22,curti22,arellanocordova22,katz23,taylor22,brinchmann22}}, despite the independent approaches to flux calibration and spectral extraction used in each work. {ID~4590 has an estimated electron temperature that exceeds the maximum value ($\log(T_{\rm e}/{\rm K}) \simeq 4.3$) 
used in the calibration of \citet{perezmontero21} and so its low metallicity represents a (modest) extrapolation of the relation.}
The sample of metal-poor $z \sim 0$ galaxies compiled by \citet{nakajima22} similarly lacks analogs to the high $T_e$ (and high-ionization line ratios) measured for the $z>5$ galaxies.

\begin{deluxetable}{lrc|cc}[t]
\tablenum{3}
\tablecolumns{5}
\tablecaption{Electron Temperature and Inferred Metallicity \label{tbl:teffz}}
\tablehead{
\colhead{ID} & \colhead{Redshift} & \colhead{$\log(M_\star/M_\odot)$} & \colhead{$\log(T_{\rm e}/{\rm K})$} & \colhead{log(O/H)+12}
}
\startdata
 4590 & 8.4957 & $7.10^{+0.14}_{-0.12}$ & $4.37\pm0.07$ & $7.49\pm0.26$ \\
 5144 & 6.3792 & $7.39^{+0.04}_{-0.03}$ & $4.18\pm0.04$ & $7.87\pm0.17$ \\
 6355 & 7.6651 & $8.23^{+0.08}_{-0.09}$ & $4.09\pm0.05$ & $8.10\pm0.17$ \\
10612 & 7.6597 & $7.72^{+0.06}_{-0.05}$ & $4.23\pm0.04$ & $7.75\pm0.19$
\enddata
\caption{{Stellar masses are from \citet{carnall22}. Error bars indicate 1$\sigma$ uncertainties, and for metallicity include the calibration uncertainty associated with the $T_e$-based relationship.}}
\end{deluxetable}

\vspace{-0.45cm}
It is important to note that the electron temperature measured in this fashion is associated with the portion of the ISM emitting the \OIII\ lines and may not be representative of the broader gas conditions. Significant gradients in density and/or ionization in the ISM may lead to a mix of high- and low-ionization regions and the \OIII\ electron temperature probes only the former. {Our use of the \citet{perezmontero21} metallicity relationship also implicitly assumes that the high-redshift galaxies have the same relationship between metallicity and \OIII\ electron temperature as the calibration sample of high-ionization galaxies at low redshift.} Nonetheless, the {$T_e$-based} metallicity estimates presented in Table \ref{tbl:teffz} have excellent agreement with the low metallicities implied from the comparison to the MAPPINGS models in Figures \ref{fig:ohno} and \ref{fig:o3comp}.

\begin{figure*}[t]
\centering
\epsscale{1.1}
\plotone{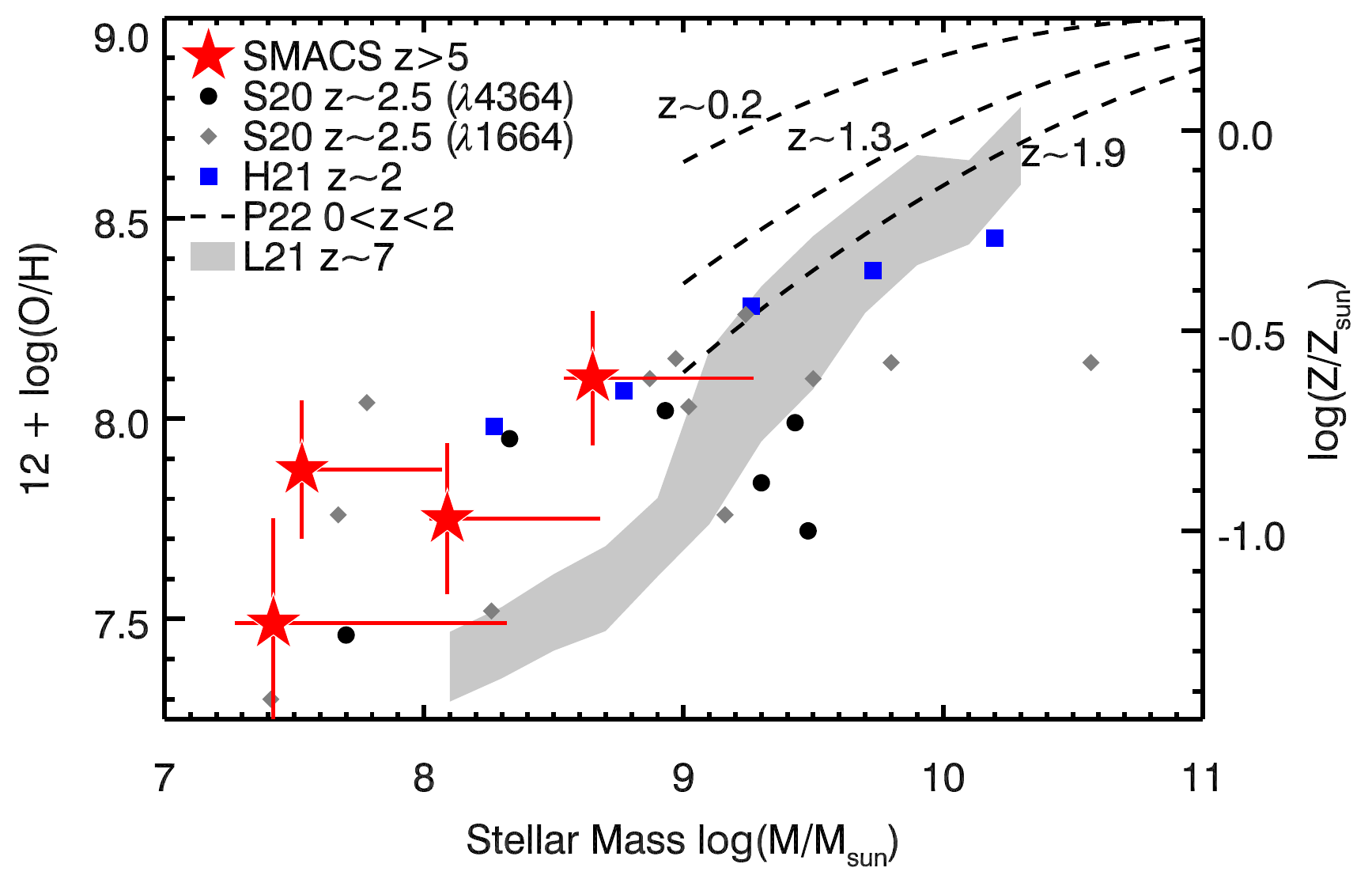}
\caption{The mass-metallicity relationship of the $z>6$ galaxies, shown as large red stars. Metallicity is estimated using the $\OIII\lambda4364/\Hg$ and $\OIII\lambda(4960+5008)/\Hb$ ratios as described in Section~4.3, and stellar masses are from \citet{carnall22}. Dashed lines indicate parametric mass-metallicity relations at $z \sim 0.2$, $z \sim 1.3$, and $z \sim 1.9$ from \citet{papovich22}.
Also shown are $z \sim 2.5$ comparison samples of galaxies with $\OIII\lambda4364$ (black points) and $\OIII\lambda1663$ (gray diamonds) ``direct'' metallicities from \citet{sanders20}, {as well as R23 metallicities} from stacked \hst/WFC3 grism measurements (blue squares) from \citet{henry21}. The gray shaded region indicates the 16-84\% distribution of galaxies at $z=7$ in the FLARES simulations \citep{lovell21}.
\label{fig:massmetallicity}}
\end{figure*}

Figure \ref{fig:massmetallicity} presents the mass-metallicity relation for the $z>6$ galaxies, using our {$T_e$-based} metallicity estimates from Table~\ref{tbl:teffz}. {The stellar masses are from \citet{carnall22} and are estimated from spectral energy distribution with a \citet{kroupa01} initial mass function, \citet{BC03} stellar population models, and nebular emission computed using \texttt{CLOUDY} \citep{ferland17}. \citet{curti22} and \citet{schaerer22} estimated significantly larger stellar masses for three of these galaxies. In order to reflect the potential systematic uncertainties, we add 0.5~dex upper error bars for the stellar masses in Figure \ref{fig:massmetallicity}.}

We compare our $z>6$ mass-metallicity properties to {a lower-redshift comparison sample} of ``direct'' 
metallicity estimates from the \citet{sanders20} compilation of $\OIII\lambda4364$ (black points) and $\OIII\lambda1663$ (gray diamonds) galaxies at $z \sim 2.5$. {The Figure also includes strong-line metallicities (using a Bayesian approach to R23) from} stacked \hst/WFC3 grism measurements of $z\sim 2$ galaxies (blue squares) from \citet{henry21}, {as well as} parametric mass-metallicity relations at $z \sim 0.2$, $z \sim 1.3$, $z \sim 1.9$ (dashed lines) determined by \citet{papovich22}.
We additionally compare our $6<z<9$ galaxies with mass-metallicity predictions of galaxies at $z=7$ in the FLARES simulations \citep{lovell21}.

At fixed stellar mass, the metallicities of our $z>6$ galaxies are generally consistent with the metallicities of auroral-\OIII-selected galaxies {of similar ($\log(M_\star/M_\odot)<9$) stellar mass} at $z \sim 2.5$.
The $z>6$ galaxies are also consistent with the stacked 
$z \sim 2$ {strong-line metallicity} measurements of \citet{henry21} {and with a low-mass extrapolation of the $z \sim 1.9$ mass-metallicity relationship of \citet{papovich22}}.
However the large stellar mass uncertainties of our $z>6$ galaxies 
mean that they are also broadly consistent with slightly lower metallicities at fixed stellar mass compared to the lower-redshift samples.

The broad distribution of mass and metallicity among the $z>6$ sample is also notable. The galaxy with the highest metallicity ($12+\log({\rm O/H})=8.1$ for ID 6355) has very bright emission lines, implying a very high star formation rate. This galaxy also has a clumpy and extended morphology that is suggestive of a merger \citep{carnall22}. The other three $z>6$ galaxies all have lower metallicities and lower stellar masses, with ID~4590 at $z=8.5$ having the lowest metallicity and lowest stellar mass of the sample.
It is interesting to measure such diversity in chemical enrichment and mass assembly in the early Universe among our limited sample of $z>6$ galaxies.

\section{Conclusions}

We use \jwst/NIRSpec spectroscopy from the SMACS~0723 Early Release Observations to study the physical conditions of the interstellar medium in five galaxies at $z>5$. We identify several caveats in the current {(v1.8.2)} reduction pipeline, including {uncertain absolute} flux calibration, too-wide spectral extractions, and underestimated uncertainties. We mitigate these issues using 
a custom spectral extraction, as described in detail in Section 2.2. We caution against uses of NIRSpec observations which require \textit{absolute} spectrophotometry or accurate continuum detection, such as equivalent widths and comparisons of lines widely separated in wavelength, especially if using the standard {(v1.8.2)} pipeline products. However we find that the \textit{relative} flux calibration is reliable, as determined by the stability of measured line ratios in different visits {and by a measured $\OIII\lambda5008/\OIII\lambda4960=3$ for all galaxies}.

We measure the ratios of rest-frame optical emission lines that are closely separated in wavelength, including $\NeIII\lambda3870/\OII\lambda3728$, $\OIII\lambda4364/\Hg$, and $\OIII\lambda5008/\Hb$. Compared to lower-redshift {($z \sim 3$)} galaxies, the $z>5$ galaxies have similar $\OIII\lambda5008/\Hb$, {similar $\OIII\lambda4364/\Hg$, and $\sim$0.5~dex higher $\NeIII\lambda3870/\OII\lambda3728$}. The $z>5$ emission-line ratios are generally well-described by MAPPING~V photoionization models for an ISM that has very high ionization ($\log(Q) \simeq 8-9$), very high pressure ($\log(P/k) \simeq 8-9$), and low metallicity ($Z/Z_\odot \lesssim 0.2$).

The $\OIII\lambda4364/\Hg$ and $\OIII\lambda5008/\Hb$ emission-line ratios indicate very high electron temperatures of $4.1<\log(T_e/{\rm K})<4.4$ in the {four $z>6$ galaxies}. We use these electron temperatures to estimate {$T_e$-based} nebular metallicities of {$7.5<12+\log({\rm O/H})<8.1$ ($Z/Z_\odot \lesssim 0.2$)} in the 4 $z>6$ galaxies.
Using stellar masses published in other work, we present a mass-metallicity diagram that compares the $z>6$ galaxies with lower-redshift samples and with theoretical simulations. The $z>6$ metallicities are broadly consistent with $z \sim 2$ galaxies of similar stellar mass, although our interpretation is limited by highly uncertain stellar masses.

These measurements demonstrate the impressive capability of \jwst\ spectroscopy for understanding the physical conditions of the gas in galaxies at cosmic dawn. We look forward to upcoming \jwst\ observations of larger samples of galaxies that will further probe the assembly and chemical enrichment of the first galaxies in the Universe.

\begin{acknowledgments}

The authors are enormously grateful to all of the people who designed, built, launched, deployed, and commissioned the \textit{James Webb Space Telescope}. The observations from this spacecraft are incredible. Seriously: much wow.

This work is based on observations made with the NASA/ESA/CSA James Webb Space Telescope. The data were obtained from the Mikulski Archive for Space Telescopes at the Space Telescope Science Institute, which is operated by the Association of Universities for Research in Astronomy, Inc., under NASA contract NAS 5-03127 for JWST. These observations are associated with program \#2736. The authors acknowledge the ERO team for developing their observing program with a zero-exclusive-access period.

The CEERS team thanks Pierre Ferruit and the NIRSpec GTO team for providing NIRSpec IPS simulated data and for general good counsel, and the STScI NIRSpec instrument team for extensive assistance regarding the \jwst\ Pipeline and data simulations. We are grateful to the referee for constructive and thoughtful comments that improved the manuscript.

This work is supported by NASA grants JWST-ERS-01345 and JWST-AR-01721. JRT and BEB additionally acknowledge support from NSF grant CAREER-1945546. RCS appreciates support from a Giacconi Fellowship at the Space Telescope Science Institute. 
RA acknowledges support from Fondecyt Regular 1202007.
PGP-G acknowledges support from grant PGC2018-093499-B-I00 funded by MCIN/AEI/10.13039/501100011033.
AY is supported by an appointment to the NASA Postdoctoral Program (NPP) at NASA Goddard Space Flight Center, administered by Oak Ridge Associated Universities under contract with NASA.

The JWST Early Release Observations (ERO) and associated materials were developed, executed, and compiled by the ERO production team:  Hannah Braun, Claire Blome, Matthew Brown, Margaret Carruthers, Dan Coe, Joseph DePasquale, Nestor Espinoza, Macarena Garcia Marin, Karl Gordon, Alaina Henry, Leah Hustak, Andi James, Ann Jenkins, Anton Koekemoer, Stephanie LaMassa, David Law, Alexandra Lockwood, Amaya Moro-Martin, Susan Mullally, Alyssa Pagan, Dani Player, Klaus Pontoppidan, Charles Proffitt, Christine Pulliam, Leah Ramsay, Swara Ravindranath, Neill Reid, Massimo Robberto, Elena Sabbi, Leonardo Ubeda. The EROs were also made possible by the foundational efforts and support from the JWST instruments, STScI planning and scheduling, and Data Management teams.

\end{acknowledgments}

\bibliography{refs}{}

\end{document}